# Long-range optical pulling force device based on vortex beams and transformation optics

Shahin Firuzi[*] and Shengping Gong[†]
*School of Aerospace Engineering, Tsinghua University, 100084, Beijing, China*

In this work a method for generating a long-range optical pulling force is presented which is realized by utilizing a vortex beam and a device designed based on transformation optics through conformal mapping. The device works by transforming an input perfect vortex beam into an almost non-paraxial plane wave, and generates a pulling force by maximizing the forward momentum of the beam. A three-dimensional full-wave analysis of the device is performed and the optical force is computed by the Maxwell stress tensor method. The simulation results show a very good agreement with the theoretical calculations.

## I. INTRODUCTION

The optical forces including conservative gradient forces which are the case in optical traps [1], and optical conveyors [2], as well as non-conservative forces applied to an object through radiation pressure [3], are caused by the interaction of electromagnetic (EM) field with matter. The EM field naturally applies a pushing force to the object in the direction of propagation, which has been the foundation of operation of solar sails and laser-pushed sails as in-space propulsion [4]. However, the negative radiation pressure or optical pulling force has been also realized [5]. In the case of a non-conservative optical force, according to the law of conservation of momentum, the fundamental requirement of realizing a negative radiation pressure (pulling force) is to increase the momentum of the scattered EM field in the direction of the propagation [6–8]. This condition can be achieved either by modifying the environment [9–16], utilizing the properties of the objects [17–20], applying specifically generated beams [6, 21–26], or by employing a combination of these conditions [8, 19, 20].

The environment can be modified by utilizing negative refractive index media which results in a negative radiation pressure applied to the particles [10, 11]. It's been also shown that the momentum of light can be increased by scattering into a medium with higher refractive index, which can be utilized to apply a pulling force to a non-absorbing arbitrary dielectric object [9]. However, a pulling force can also be applied to light-absorbing particles in a gaseous medium through photophoretic forces [12]. The concepts of using hyperbolic metamaterials [13] and metasurfaces [14], as well as photonic crystals [15, 16] as the environment around the object, have been also proposed to modify the interaction of the EM field with the object which can be used to generate a pulling force.

Increasing the forward momentum of the scattered field, can be also achieved by using an already excited gain medium [17, 18]. However, the gain medium needs to be in a populated state to be able to generate a pulling force, and by considering a single source for populating and pulling the object, according to the law of conservation of momentum, it does not generate any total negative force. It has been shown that optical pulling forces can be also induced in dipole and multipole chiral particles through coupling of angular momentum to linear momentum [19, 20]. However, structure fields are required to realize such negative forces [3].

Structured fields can be utilized as non-paraxial incident beams (with a non-zero Poynting vector component perpendicular to the propagation direction) which inputs less momentum in the direction of propagation. Increasing the forward momentum of such beams through scattering by an object (decreasing the non-paraxiality of the beam), applies a force to the object which is opposite to the propagation direction [6–8, 21, 22, 26]. The proposed methods operate by transforming a radially-directed Poynting vector component (generated by interference of multiple plane waves), to the propagation direction. However, the magnitude of this radial component such as in Bessel beams, depends on the size of the aperture relative to the distance to the object (focal distance), and vanishes for long-distance targets. In addition to the non-paraxial beams, several unconventional beams have been also shown to trap microscopic particles and pull them against the propagation direction, such as super-oscillating rotating beams [23], and solenoid beam [24, 25].

Most of the methods have been mentioned above are either limited by the environment [9–16], the distance to the object [6, 8, 21, 22, 26], or the size of the object [6, 7, 22, 23, 25]. However, the idea of manipulating large objects at long distances by a single source, which can be used as a means of propulsion, such as pulling a laser-driven sail in space towards the source of light, is intensely appealing and seems to lie in the realm of science fiction. In this paper, we demonstrate how by designing both the beam and the device, such a desirable long-range pulling force can be applied to an arbitrary-sized device by means of non-conservative optical forces (i.e. negative radiation pressure). We show that such a negative radiation pressure can be generated by using a single non-paraxial beam and a reflection-less device designed based on transformation optics [27, 28]. The negative radiation pressure is generated by increasing the forward momentum of the beam through transforming it into a paraxial beam. However, as we mentioned earlier, the non-paraxiality caused by a radially-directed Poynting vector component, vanishes for large focal distances. However, a non-paraxiality can be achieved by an azimuthally-directed Poynting vector component which is independent of the focal distance. Therefore, by employing an azimuthally-directed Poynting vector component, which is the case for vortex beams, a long-range optical pulling force beam may be achieved.

---

[*] xiah16@mails.tsinghua.edu.cn
[†] gongsp@tsinghua.edu.cn

## II. THEORETICAL BASIS

The Poynting vector of a vortex beam carrying orbital angular momentum (OAM) [29], can be understood by imagining the photons move along the propagation axis while rotating around it. The Poynting vector makes an angle with the propagation axis which theoretically can be designed by a desired value. We consider a perfect vortex beam [30], which has a ring diameter independent of its topological charge. Although other types of beam carrying OAM can be considered as a long-range non-paraxial beam, considering a perfect vortex beam enables the design of an arbitrary-sized device independent of the chosen topological charge. The field amplitude of a perfect vortex beam at the beam waist, in the cylindrical coordinates ($\rho,\phi,z$), can be expressed as [30]

$$E(\rho,\phi) = A_0 \left(-1\right)^l \exp(-\frac{\rho^2 + r_r^2}{w_0^2}) I_l \left(\frac{2r_r\rho}{w_0^2}\right) \exp(il\phi + ikz), \quad (1)$$

where $A_o$ is a tunable constant, $l$ is the topological charge, $k$ is the axial wave vector, and $I_l$ is an $l$th order modified Bessel function of first kind. This beam has a ring width and radius of $2w_o$ and $r_r$, respectively. The time-averaged Poynting vector $\boldsymbol{S}$ makes a $\theta_i$=tan$^{-1}$($l/\rho k$) angle with the propagation axis. The direction of flow of the energy can be expressed by a unit vector given by cylindrical unit vectors ($\hat{e}_\rho, \hat{e}_\phi, \hat{e}_z$) as

$$\hat{s} = \frac{l}{\sqrt{l^2 + k^2\rho^2}} \hat{e}_\phi + \frac{k\rho}{\sqrt{l^2 + k^2\rho^2}} \hat{e}_z. \quad (2)$$

Therefore, the time-averaged Poynting vector of the beam in vacuum can be given as

$$\boldsymbol{S} = \frac{c\varepsilon_0}{2} \|\boldsymbol{E}(\rho,\phi)\|^2 \hat{s}, \quad (3)$$

where $c$ is the speed of light in vacuum, $\varepsilon_o$ is the vacuum permittivity, and the notation $\|\cdot\|$ represents the magnitude of a function. Therefore, the momentum flux vector of the beam can be given by

$$\boldsymbol{p}_{beam} = \frac{1}{c} \int_0^{2\pi} \int_{r_r - w_0}^{r_r + w_0} \boldsymbol{S} \rho \, d\rho \, d\phi. \quad (4)$$

From Eq. (3), it can be seen that the magnitude of $\boldsymbol{S}$ is independent of its direction. Therefore, the direction of $\boldsymbol{S}$, which is defined by $\hat{s}$, can be modified by a passive device without necessarily changing the energy content of the beam. Any changes of this kind will alter the direction of $\boldsymbol{p}_{beam}$ without changing its magnitude. By considering a system consisting of the beam and the device, according to the law of conservation of momentum, the changes in the momentum flux vector of the beam and the device, are equal and in the opposite direction (i.e. $\Delta \boldsymbol{p}_{device}$= -$\Delta \boldsymbol{p}_{beam}$). According to Eq. (2), decreasing the $\hat{e}_\phi$ component of the Poynting vector (decreasing the non-paraxiality of the beam), increases its $\hat{e}_z$ component, which consequently increases $\boldsymbol{p}_{beam}$ along the propagation direction. This change, therefore, applies a force to the device in the direction opposite to the propagation direction of light (i.e. optical pulling force). By considering a device composed of non-absorbing materials (e.g. SiO$_2$) in the operating wavelengths, the optical pulling force can be given by

$$F_z = \frac{\varepsilon_0}{2} \int_0^{2\pi} \int_{r_r - w_0}^{r_r + w_0} \|\boldsymbol{E}(\rho,\phi)\|^2 \left(\hat{s}_z^i - \hat{s}_z^o\right) \rho \, d\rho \, d\phi, \quad (5)$$

where the subscripts $i$ and $o$ correspond to the device's input and output beams, respectively. The maximum magnitude of the optical pulling force can be achieved by transforming the vortex input beam into a paraxial output beam (i.e. $\hat{s}_\phi^o = 0, \hat{s}_z^o = 1$). Such a device transforms the perfect vortex beam into an annular shaped Gaussian beam, and consequently maximizes the momentum of the beam along its propagation direction. In this case, by assuming $r_r >> w_o$, and by having the total power ($P_{total}$) of the vortex beam, the maximum pulling force for a given $\theta_i$ can be expressed as

$$F_z^{max} = \frac{P_{total}}{c} \left(\cos(\theta_i) - 1\right), \quad (6)$$

where $\theta_i$=tan$^{-1}$($l/\rho k$). From Eq. (6), it can be seen that by increasing the $\theta_i$ angle (increasing $l/\rho k$), the magnitude of the pulling force per unit power increases, and for large values of $l/\rho k$, it gets close to the magnitude of the maximum possible optical force by absorption (i.e. $P_{total}/c$).

The beam also applies a torque to the device, which by assuming $r_r >> w_o$, can be approximated as

$$T_z = \frac{\varepsilon_0 r_r}{2} \int_0^{2\pi} \int_{r_r - w_0}^{r_r + w_0} \|\boldsymbol{E}(\rho,\phi)\|^2 \left(\hat{s}_\phi^i - \hat{s}_\phi^o\right) \rho \, d\rho \, d\phi. \quad (7)$$

However, by having the total power ($P_{total}$) of the vortex beam, the maximum torque for a given $\theta_i$ can be

expressed in a simplified form as

$$T_z^{\max} = \frac{P_{\text{total}}}{c} r_r \sin(\theta_i). \tag{8}$$

## III. DESIGN OF THE DEVICE

The device used for generating the optical pulling force, needs to guide the EM wave in a reflection-less fashion to avoid the back scattering of the beam which may reduce the pulling force. Transformation optics (TO) [27, 28] is a great candidate for the design of such a device. TO techniques have been utilized to design optical devices which can control the behavior of electromagnetic field, such as waveguides and wave shifters\splitters\combiners, invisibility and illusion devices, flat lenses, low-profile highly directive antennas, and many other remarkable devices [31, 32]. TO devices have been also used to generate vortex beams [33, 34], the design of which, however, needs anisotropic materials [34]. For such designs, due to azimuthally varying refractive index at the input and output ports of the device [33, 34], index-matching can't be applied, which may result in reflections and backscattering of the light on the input/output boundaries.

In this section we show that a device designed based on TO using conformal mapping [28], can be utilized to modify the Poynting vector of a perfect vortex beam in a desired manner, while minimizing the reflections on the input/output boundaries. The reflections on the boundaries are minimized using an index matched layer, by matching the refractive index of the device and the surrounding environment on the input/output boundaries. A TO device can be realized by using metamaterials. However, metal-dielectric metamaterials have large losses in optical wavelengths, which may influence the generation of the pulling force by decreasing the forward momentum of the output beam. Therefore, implementing a low-loss all-dielectric design based on conformal mapping techniques [28, 35] is necessary.

The conformal mapping can be applied to a two-dimensional device, by solving the Laplace's equation over a certain boundary condition [35]. However, this technique has been also applied to three-dimensional devices which are identical along the third dimension [31]. From Eq. (2) it can be seen that the unit vector $\hat{s}$ depends on $\rho$, which implies that the Poynting vector of a vortex beam is not identical along $\hat{e}_\rho$. However, by considering a perfect vortex beam which its ring radius is so much larger than the ring width ($r_r >> w_o$), the dependency of $\hat{s}$ on $\rho$ can be minimized, and the device can be considered as identical along $\hat{e}_\rho$. Therefore, a two-dimensional conformal mapping can be applied to the $\hat{e}_\phi \hat{e}_z$ surface corresponding to $\rho = r_r$, and the results can then be extended along $\hat{e}_\rho$.

The desired trajectory of a photon on the $\hat{e}_\phi \hat{e}_z$ surface as controlled by the device, is shown in Fig. 1(a) by the red-and-blue curve (red lines and blue curve). It can be seen that the Poynting vector is modified along this curve and redirected towards the z axis. As shown in Fig. 1(b), a two-dimensional waveguide is formed by cutting the cylindrical $\hat{e}_\phi \hat{e}_z$ surface along this curve, and spreading it on a plane. The conformal mapping is applied to the green region of this waveguide shown in Fig. 1(b), and the blue regions are the index matching layers. Two conformal components of $\zeta$ and $\eta$ which can be given by solving the Laplace's Equation, correspond to the wavefront and propagation path of photons, respectively. However, according to the Cauchy-Riemann equations [35], the refractive index of the two-dimensional waveguide can be given merely by the partial derivatives of one of the conformal components as [35]

$$n = n_{\text{env}} \sqrt{\zeta_x^2 + \zeta_y^2}, \tag{9}$$

where $n_{\text{env}}$ is the refractive index of the surrounding environment, and the subscript $x$ shows the partial derivative with respect to $x$. The conformal mapping is applied to the green region of Fig. 1(b) by solving the Laplace's equation considering the Dirichlet boundary conditions on $\Gamma_1$ and $\Gamma_5$ lines, and the Neumann boundary conditions on $\Gamma_2$, $\Gamma_3$, $\Gamma_4$, $\Gamma_6$, $\Gamma_7$, and $\Gamma_8$ curves as

$$\zeta|_{\Gamma_1} = 0, \zeta|_{\Gamma_5} = a,$$
$$\text{and } N_{\Gamma_i} \cdot \nabla \zeta|_{\Gamma_i} = 0 \text{ for } i = 2:4, 6:8, \tag{10}$$

where $N_{\Gamma_i}$ is the unit normal vector to the curve $\Gamma_i$. In order to guarantee the all-dielectric structure of the waveguide, the constant $a$ is defined in such a way to realize $n_{\min} \geq 1$, where $n_{\min}$ is the minimum refractive index of the waveguide.

In order to apply index matching to the input/output ports of the device, it is necessary to achieve a nearly uniform refractive index on the $\Gamma_1$ and $\Gamma_5$ boundaries. Since the alteration of the propagation path of the beam in the waveguide (Fig. 1(b)) occurs mostly in the curved region between $\Gamma_3$ and $\Gamma_7$, the refractive index has a large variation in this region. However, it does not become uniform instantly outside this region, and by increasing the length of $\Gamma_2$, $\Gamma_4$, $\Gamma_6$, and $\Gamma_8$, the refractive index becomes more uniform on $\Gamma_1$ and $\Gamma_5$. Therefore, the length of the device needs to be long enough to be able to get a desirably uniform refractive index at the ports of the waveguide. This may result in a very long device compared to its width, which increases the mass-to-illuminated-surface ratio of the device, and consequently reduces the resultant acceleration and the effectiveness of the optical pulling force. However, by dividing the $\hat{e}_\phi \hat{e}_z$ surface of the device into multiple ($m$ number of) waveguides, the length of the waveguides can be reduced, which results in a shorter and lighter device. Therefore, by keeping the size of every waveguide ($2\pi r_r/m$) in the geometric optics regime, the length of the device can be reduced to as small as tens of the wavelength of the beam. In this way, designing thin and light devices will become possible. At the ports of the waveguide an index matching layer (Fig. 1(b), blue region) is applied to minimize the reflections. The resultant gradient refractive index can be implemented using all-dielectric metamaterials by graded photonic crystals [36], or by drilling radial holes into the dielectric materials [37, 38]. However, by considering the wavelength of the beam, the diameter of the holes needs to be in the range of tens to hundreds of nanometers.

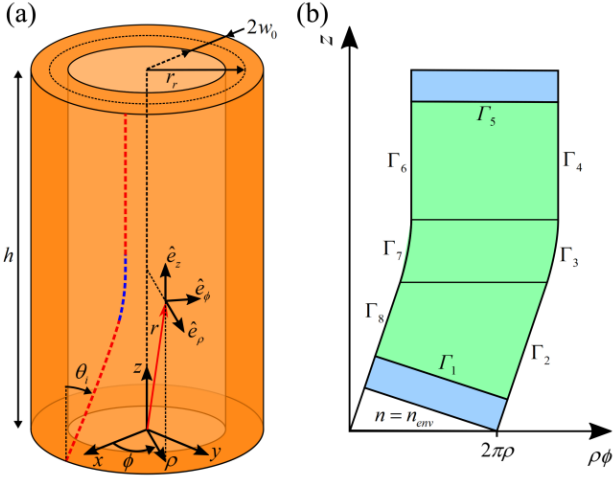

FIG. 1. The structure of the device and the waveguide constructing the device. (a) Shows the annular structure of the device along with the cylindrical reference frame. The red vector represents the location vector of a point of the device from the origin. (b) Shows the structure of the two-dimensional waveguide formed by cutting the cylinder's surface and spreading it over a plane.

## IV. SIMULATION AND RESULTS

In order to prove the existence of a negative radiation pressure, performing a three-dimensional full-wave analysis of the device is necessary. Since the device is divided into multiple identical waveguides with completely identical conditions, performing the simulation on a single waveguide is sufficient to show the performance of the whole device. The device is considered to be an annular cylinder divided into 90 waveguides ($m$=90). The source is a perfect vortex beam with $r_r$=170µm, $w_o$=9.5µm, $l$=180, a wavelength of $\lambda$=1.064µm, and a total power of 90 watts, which results in 1 watt power on a single waveguide. However, for large topological charges, Eq. (1) results in slightly different ring radius coinciding with the maximum intensity of the beam. Therefore, the ring radius of the device is calculated as $r_r$=176.047µm. The ring width and the length of the device are $2w_o$=19µm and $h$=70µm, respectively.

The conformal mapping is performed by solving the Laplace's equation numerically, and the resultant two-dimensional refractive index is extended along the third dimension ($\hat{e}_\rho$) to generate the three-dimensional refractive index. The performance of a three-dimensional waveguide is simulated by finite-difference time-domain (FDTD) simulations with commercial software (Lumerical FDTD Solutions), and the electromagnetic fields are computed. The optical forces are computed using Maxwell stress tensor method, by a surface integral of the time-averaged Maxwell stress tensor over the three-dimensional waveguide's surface.

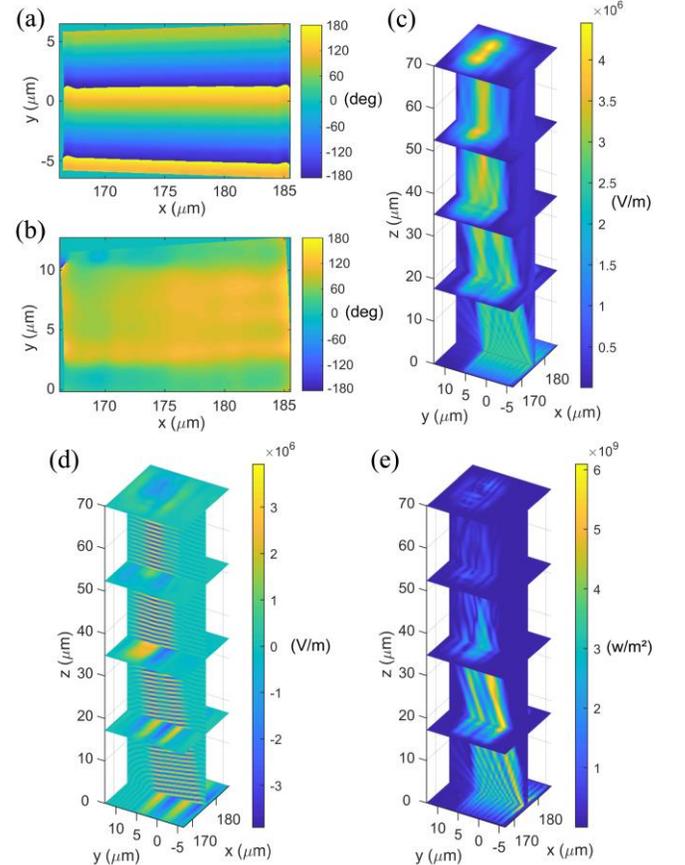

FIG. 2. Results of the FDTD simulation. (a) Shows the phase of the input perfect vortex beam. (b) presents the phase of the output beam. The output phase is nearly uniform, which indicates that the output beam is almost a plane wave. The magnitude of the electric field and its $x$ component are shown in (c) and (d), respectively. (e) Shows the azimuthal component of the time-averaged Poynting vector, which vanishes along the trajectory.

The phase of the perfect vortex beam at the entrance port of the device is shown in Fig. 2(a), and the phase of the output beam is shown in Fig. 2(b). It can be seen that the phase of the output beam is almost uniform, which indicates that the wavefront is almost flat and perpendicular to the propagation direction (paraxial beam). The distribution of the magnitude of the electric field is shown in Fig. 2(c). It can be seen that, as the wave propagates in the device, the direction of the power flow changes towards the $z$ axis. From Fig. 2(d) it can be seen that the $x$ component of electric field ($E_x$) also follows the same path. It also shows that the structure of the electric field is maintained along the propagation in the device. The time-averaged azimuthal component of the Poynting vector which is shown in Fig. 2(e), vanishes with the propagation of the beam in the device. It shows that the direction of the energy flow is altered, and the azimuthal component of the Poynting vector is transferred to the axial Poynting vector component.

Maxwell stress tensor results show a pulling force of $F_z^{sim}$=-5.0616×10$^{-11}$N is applied to the waveguide. The forces along $x$ and $y$ axes are computed as $F_x^{sim}$=-2.3338×10$^{-11}$N and $F_y^{sim}$=5.9191×10$^{-10}$N, respectively. The torque generated by $F_x^{sim}$ and $F_y^{sim}$ can be estimated as $T_z^{sim}$=1.0429×10$^{-13}$N.m. The theoretical optical forces applied to the device is calculated by Eq. (5) as $F_z^{theory}$=-5.2915×10$^{-11}$N, $F_x^{theory}$=-2.1474×10$^{-11}$N, and $F_y^{theory}$=6.1493×10$^{-10}$N, which results in a theoretical torque of $T_z^{theory}$=1.046×10$^{-13}$N.m. It can be seen that there is a good agreement between the theoretical and simulation results. However, the theoretical optical forces are calculated under ideal conditions, which the reflections on boundaries and the imperfections of the gradient refractive index haven't been considered. Therefore, these imperfections result in small differences between the simulation and theoretical results.

## V. CONCLUSION

This paper demonstrated a device for generation of non-conservative optical pulling force (negative radiation pressure), by transforming a vortex beam into an almost paraxial beam. Since the non-paraxiality of a vortex beam is caused by the azimuthal component of the Poynting vector which is independent of the focal distance, the pulling force can be realized at very long distances. The device which is designed based on transformation optics, modifies the wavefront of an input perfect vortex beam to an annular plane wave, and thereby maximizes the forward momentum of the output beam, which consequently generates an optical pulling force. The magnitude of the pulling force per unit power for a certain wavelength has a direct relationship with the topological-charge-to-ring-radius ratio. Accordingly, for the large values of this ratio, the magnitude of the pulling force per unit power approaches the value of the maximum pushing force per unit power by absorption (i.e. half of the optical force due to complete specular reflection). The presented device may have applications in long-range optical manipulation of large objects such as in-space photonic propulsion.

## ACKNOWLEDGEMENTS

This work is supported by the National Natural Science Foundation of China (NSFC) under Grant Nos. 11772167 and 11822205. We would like to express our deep gratitude to Professor Benfeng Bai for his valuable and constructive suggestions which helped us to improve the quality of this work.